\documentclass[dvips]{kluwer}    
\usepackage[dvips]{graphicx}
\begin{document}
\begin{article}
\begin{opening}
\title{
Current Helicity and Twist as Two \\
 Indicators of The Mirror Asymmetry of
Solar Magnetic Fields
}
\author{D. \surname{
Sokoloff
}}
            \institute{National Astronomical
Observatories, Chinese Academy of Sciences, Beijing 100012, China and Department of
Physics, Moscow State University, Moscow 119992, Russia}
\author{H. \surname{Zhang}}
            \institute{National Astronomical
Observatories, Chinese Academy of Sciences, Beijing 100012, China}
\author{K.M. \surname{Kuzanyan}}
            \institute{National Astronomical
Observatories, Chinese Academy of Sciences, Beijing 100012, China, IZMIRAN, Troitsk,
Moscow Region 142190, Russia and School of Mathematics, University of Leeds,
Leeds LS2 9JT, UK}
\author{V.N. \surname{Obridko}}
\institute{IZMIRAN, Troitsk, Moscow Region 142190, Russia}
\author{D.N. \surname{Tomin}, V.N. \surname{Tutubalin}}
            \institute{Department of Mechanics and Mathematics, Moscow State University,
  Moscow 119992, Russia}
\runningtitle{Current helicity and twist}

\date{(Received ..... 2007; accepted ....)}

\begin{abstract}
A comparison between the two tracers of magnetic field mirror
asymmetry in solar active regions, twist and current helicity, is
presented. It is shown that for individual active regions these
tracers do not possess visible similarity while averaging by
time over the solar cycle, or by latitude, reveals similarities in
their behaviour. The main property of the dataset is anti-symmetry over
the solar equator. Considering the evolution of helical properties
over the solar cycle we find signatures of a possible sign change at
the beginning of the cycle, though more systematic observational data are
required for a definite confirmation.
We discuss the role of both tracers in the
context of the solar dynamo theory.

\end{abstract}

\keywords{solar activity; solar magnetic fields}

\end{opening}

\section{Introduction}

Contemporaty astronomical observations suggest two proxies for mirror asymmetry in solar
active regions, namely, the current helicity and twist of magnetic field. Both proxies,
averaged over a suitable part of active regions, have been measured recently for
several hundred active regions over more than one solar cycle (see, {\it e.g.}, Bao and
Zhang, 1998 and references therein). Both proxies demonstrate, to some extent, an
antisymmetry with respect to the solar equator as well as a cyclic behaviour on
timescales of the solar activity cycle as traced by sunspots.

Solar magnetic field structure is complicated enough to allow many proxies for
its mirror asymmetry, which may not necessarily be proportional one to another. As is
natural to expect, current helicity and twist have no simple relation between
each other.

A detailed comparison between current helicity and twist as two proxies of the
mirror asymmetry of the solar magnetic field is interesting in the context of solar dynamo
theory;  the key driver of the solar dynamo suggested by
Parker (1955) is the  $\alpha$-effect originating in the mirror asymmetry of
solar convection and magnetic fields. Proxies of mirror asymmetry of the magnetic
field in solar active regions provide a unique observational approach for the direct
verification and observation of the $\alpha$-effect. The link between the
$\alpha$-effect and the proxies under discussion is usually given in terms of the current
helicity ({\it e.g.}, Kleeorin {\it et al.}, 2003). It would however be risky to insist that
solar dynamo theory is well-developed enough to disregard twist as an alternative
tracer of the $\alpha$-effect. A conventional mean-field description of the
$\alpha$-effect deals with quantities averaged over substantial temporal or spatial
domains rather than over an individual active region. If the similarity of current helicity
and twist as tracers of mirror asymmetry becomes more pronounced after averaging,
it means that both proxies are reasonable tracers of the $\alpha$-effect and support the
conventional concept of the solar dynamo.

In the present paper we compare current helicity and twist data. First, we study the
proxies for an individual active region to
demonstrate that the correlation between current helicity and twist is rather weak. 
Then we demonstrate that the similarity becomes much more pronounced after temporal 
or spatial averaging.

Our investigation is based on data obtained at the Huairou Solar Observing station
of the National Observatories of China (Zhang and Bao, 1998). A previous analysis of
current helicity and twist based on this data was presented by Zhang {\it et al.} (2002), see
also Kuzanyan {\it et al.} (2000). Here we use a larger data set and improve the
statistical analysis, as well as embedding the study in the context of the solar
dynamo in a more explicit form.

The paper is organized as follows. We briefly review the concepts of current helicity
and twist as they are exploited in theoretical studies and give their observational
proxies (Section~2). 
Then we describe the observational data set (Section~3), compare the
current helicity and twist for a particular active region (Section~4) and
then present the data after temporal or spatial averaging (Section~5). Section~6 contains
a more detailed analysis of the antisymmetry of the proxies with respect to the solar
equator. We discuss the consequences for solar dynamo theory from this
analysis in Section~7.

\section{Current Helicity and Twist and the $\alpha$-effect}

The conventional parametrization of the magnetic
contribution to the $\alpha$-effect ({\it e.g.}
Kleeorin and Rogachevskii, 1999) is based on the current
helicity $\chi^{\rm c} = <\!\! {\bf b \cdot j} \!\!>$, where ${\bf j} =
{\rm curl} \, {\bf b}$ is the electric current and $\bf b$ is the
(small-scale) magnetic field. Because ${\rm curl} \, {\bf b}$ is
calculated from the surface magnetic field distribution, the only
electric current component that can be derived is $({\rm curl}
\,{\bf b})_z$.  As a consequence of these restrictions, the
observable quantity is

\begin{equation}
H_{\rm c} = \langle b_z ({\rm curl} \, {\bf b})_z \rangle \; ,
\label{observ}
\end{equation}
where $x,y,z$ are local cartesian coordinates connected with a point
on the solar surface, and the $z$-axis is normal to the surface (Bao
and Zhang, 1998; see also  Abramenko {\it et al.}, 1997; Pevtsov {\it et al.},
1994). In the framework of the hypothesis of local homogeneity and
isotropy this value is $1/3$ of the current helicity $\chi^{\rm c}$.

Observations ({\it e.g.} Zhang, and Bao, 1998) provide another proxy for
the mirror asymmetry of the magnetic field, {\it i.e.} twist (Woltjer,
1958) which comes from studies of magnetic fields in the solar
atmosphere, where conductivity is high.
However, because of low-beta condition the magnetic field can be described 
as force-free.
Furthermore, according to
Maxwell's equations, the magnetic field is a Beltrami field, {\it i.e.},
${\rm curl} \, {\bf b} = \alpha_{\rm ff} {\bf b}$, where the
parameter $\alpha_{\rm ff}$ is the twist. In the solar interior,
however, the magnetic field is not a Beltrami field and twist can be
understood as $\alpha_{\rm ff} = <\!\! {\rm curl} \, {\bf b} \cdot
{\bf b}/b^2 \!\!>$. Of course, this definition does not coincide
with that of the current helicity. The observational equivalent of
the quantity $\alpha_{\rm ff}$ is the ratio $<\!\! j_z/b_z \!\!>$.
The notation $\alpha_{\rm ff}$ is generally used for twist in the
solar physics literature, though it seems rather confusing from the
viewpoint of solar dynamo theory. The details of calculation of
twist and helicity from magnetographic observational data are given
in literature ({\it e.g.} Wang {\it et al.}, 1996; Bao and Zhang, 1998; Zhang
and Bao, 1998).

\section{Observational Data}

The observational data used in our analysis were obtained at the
Huai\-rou Solar Observing station of the National Astronomical
Observatories of China. A magnetograph using the Fe{\sc i} 5324 \AA \,
spectral line determines the magnet\-ic field values at the level of
photosphere. The data are obtained using a CCD camera with $512
\times 512$ pixels over the whole magnetogram. The entire image size
is comparable with the size of an active region, which at about
$2\times 10^{8}$\,m is comparable with the depth of the solar
convective zone.

However, the observational technique allows the line-of-sight
field component $b_z$ to be determined with a much higher precision
than the transverse components ($b_x$ and $b_y$). There are a number
of other observational difficulties, such as resolving the
$180^\circ$ ambiguity in the direction of transverse field
etc. The observational technique is described in detail by Wang et
al. (1996), see also Abramenko {\it et al.} (1996).

The observations are restricted to active regions on the solar
surface and we obtain information concerning the surface magnetic
field and helicity only. Monitoring solar active regions while
they are passing near the central meridian of the solar disc enables
observers to determine the full surface magnetic field vector. The
observed magnetic field is subjected to further analysis to
determine the value of $ {\rm curl} \, {\bf b}$.

An observational programme to reveal the values of the twist and the
current helicity density over the solar surface requires a
systematic approach, both to the monitoring of magnetic fields in
active regions and to the data reduction, in order to reduce the
impact of noise. This work has been carried out by a number of
research groups ({\it e.g.}. Seehafer 1990; Pevtsov {\it et al.} 1994; Rust and
Kumar 1996; Abramenko {\it et al.} 1997; Bao and Zhang 1998; Kuzanyan {\it et al.}
2000).

The present work analyzes the two systematic datasets of active
regions. The first one consists of 422 active regions over the 10
years 1988-1996 (Bao and Zhang, 1998). It has been used for
theoretical analysis and further data reduction by Kuzanyan {\it et al.}
(2000), Zhang {\it et al.} (2002), Kleeorin {\it et al.} (2003), Zhang {\it et al.}
(2006) and Sokoloff {\it et al.} (2006).

We also analyse a dataset which covers
the three years at the beginning of the solar cycle 23, namely
1998-2000. This dataset was discussed earlier by Bao {\it et al.} (2000,
2002), and contains data for 64 active regions for which all the
helical quantities were determined. The new data are obtained using the
same technique and are processed in much the same way, as the earlier
dataset of Bao and Zhang (1998) covering the ten year period
1988-1997, see also Zhang and Bao (1998). Thereafter, following
Sokoloff {\it et al.} (2006), we merge these two sets of data and
henceforth will consider them as a single continuous dataset of 486
active regions.

Observational work is ongoing, and much more data are due to be
processed shortly ({\it e.g.}, Xu {\it et al.}, 2007).

\section{Helicity and Twist for a Particular Active Regi\-on}

We consider both observable parameters for a particular active
region as random quantities.
The standard way to represent a random quantity is as a
histogr\-am presenting the percentage of data lying in a given range. We tried this method
and found that
the result is not very informative. The point is that our data set is
small and the data are quite noisy. So we used a more advanced method
and calculated a cummulative distribution function (c.d.f.) which is much
more robust than the usual  probability distribution function.

\begin{figure}
\begin{center}
\includegraphics[height=2.1in,width=4.1in, angle=0]{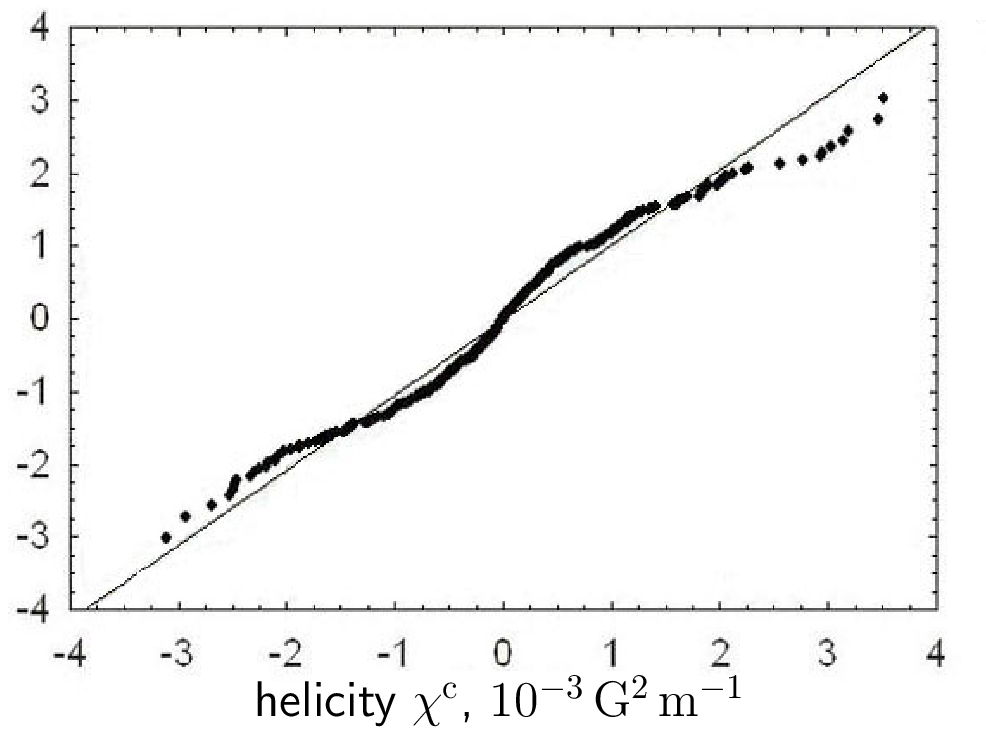}
\vbox{
\vspace{2mm}
\hbox{
\hspace{-2mm}
\includegraphics[height=2.1in,width=4.0in, angle=0]{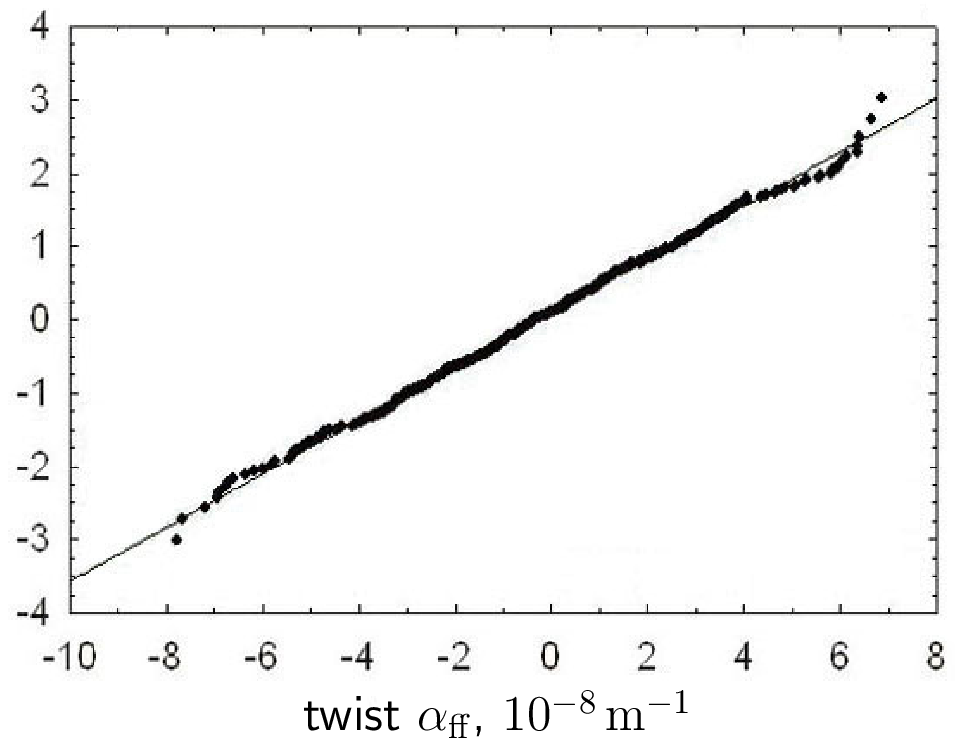}
}
}
\vspace{-2mm}
  \caption{Cumulative distribution function for the current helicity 
(upper panel) and twist (lower panel). 
The first coordinate of the point is the
value of the parameter under investigation for a particular active
region. The second coordinate is the expected value of the standard
deviations for a Gaussian quantity with the same mean and standard
deviation which gives the same probability (see in text for
details). For a Gaussian quantity, we must obtain straight lines,
which are also shown on both panels.}
\end{center}
  \label{fig1}
\end{figure}

Let our set contain $N$ active regions. For the sake of definiteness
consider a given value of  $x= \alpha_{\rm ff}$ (or $\chi^{\rm c}$
if appropriate). Let $n$ active regions have twist lower than $x$.
Then the probability for $\alpha_{\rm ff}$ to be lower than $x$ is
estimated as $P=n/N$. Let $\xi$ be a Gaussian variable with the same
mean value $\mu$ and standard deviation $\sigma$ as $\alpha_{\rm
ff}$ and $y$ the value for which the probability for   $(\xi
-\mu)/\sigma$ to be lower than $y$ is $P$. The results for various
$x$ are shown by dots in coordinates $(x,y)$ (Figure~1; upper pannel
is for twist and the lower one is for currrent helicity) and can be
compared with the c.d.f. for a Gaussian distribution shown by the
solid line. We see that the dots (twist) on the upper pannel are
substantially closer to the straight line than on the lower panel
(helicity). We conclude that twist is much closer to a Gaussian
random quantity than helicity. Note that the link between both
quantities is very nonlinear, and so at least one of the quantities
(here the helicity) has to be non-Gaussian.

\begin{figure}
\hbox{
\hspace{1mm}
  \includegraphics[height=2.5in,width=4.0in, angle=0]{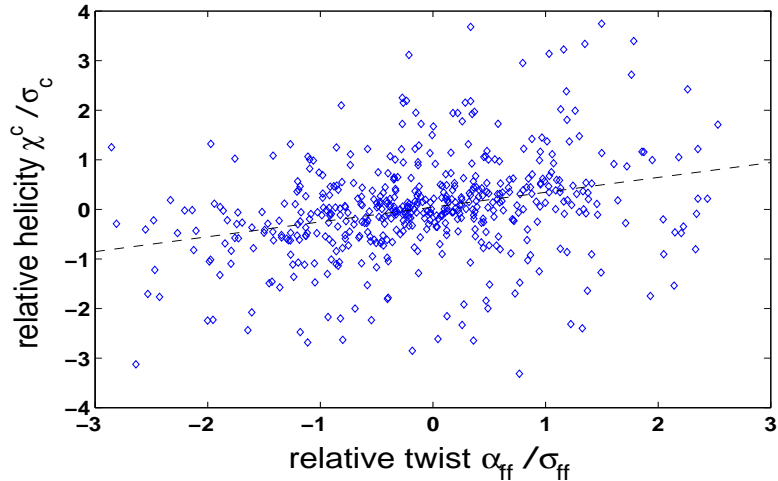}
}
\caption{Scatter diagram for the twist and helicity normalized to
the corresp\-onding standard deviations. A correlation between the
twist and helicity means that the distribution of points will form
an ellipsoid, whose inclination to the coordinate axes provides the
correlation coefficient. In the case under discussion, the
correlation is seen to be weak, though a trend (shown by a dashed
line) is noticeable.} \label{fig2}
\end{figure}

Next, the twist and helicity of the observational data normalized to
their means ($\mu_{\rm ff}$ and $\mu_c$, correspondingly) and
standard deviations ($\sigma_{\rm ff}$ and $\sigma_c$,
correspondingly) are shown on a scatter diagram (Figure~2). If the
statistical dependence between the twist and helicity is strong, the
cluster of points obtained must form an elongated ellipse. In our
case the axes of the ellipse are nearly parallel to the coordinate
axes, and a weak correlation can be described by the relation
$(\chi^{\rm c} - \mu_{\rm c})/ \sigma_{\rm c} = 0.006 +
0.1(\alpha_{\rm ff} -\mu_{\rm ff})/\sigma_{\rm ff}$ obtained by the
least square method. The resulting correlation may be formally
significant however this is difficult to confirm because of the
non-Gaussian distribution of helicity. Anyway, the correlation
revealed appears to be robust with respect to discarding data that
strongly deviates from the mean. Note, however, that theoretical
considerations based on such weak correlations are not very
reliable.

\begin{figure}
%
\includegraphics[height=2.5in,width=4.5in, angle=0]{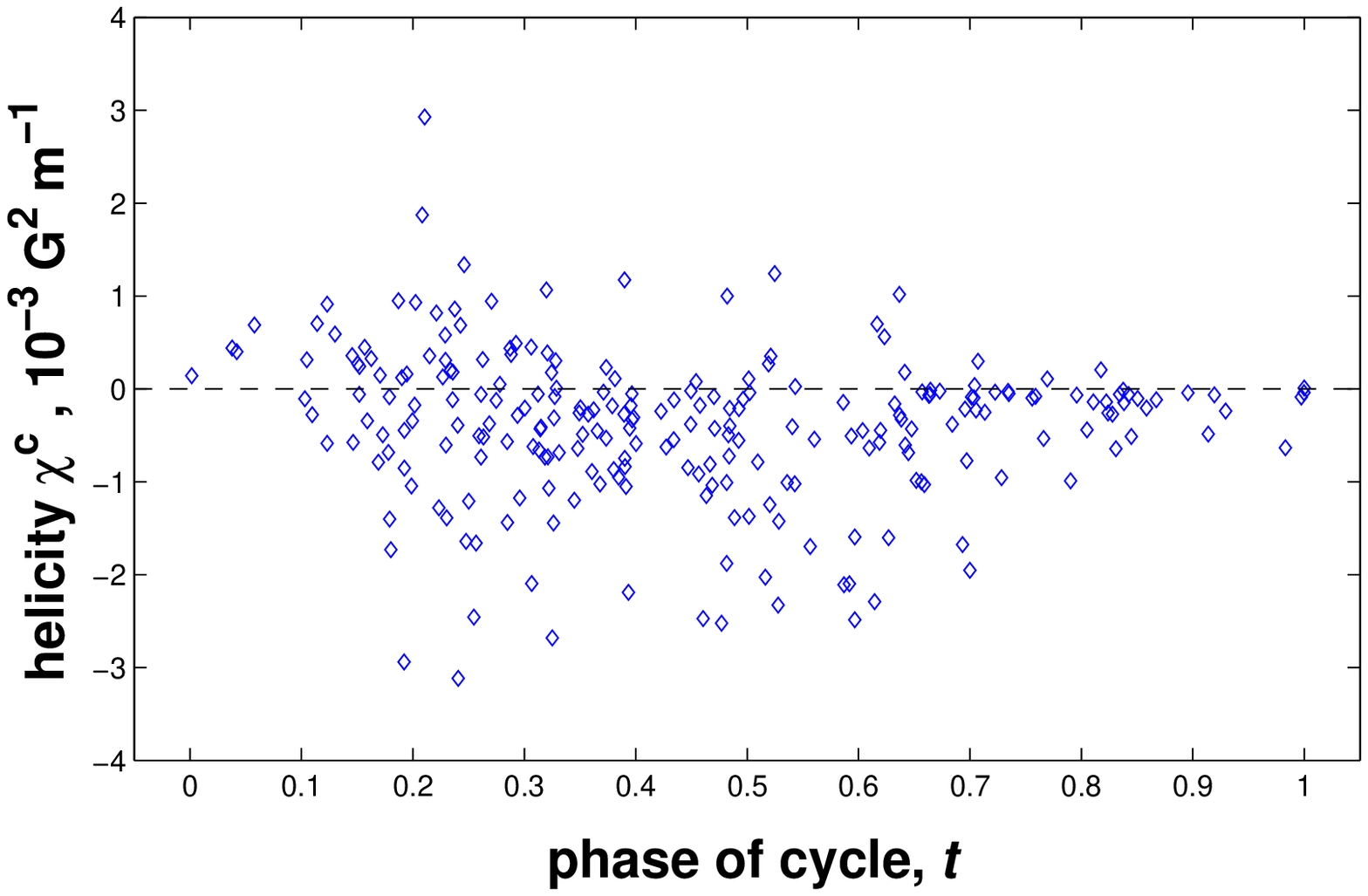}
\vspace{10mm}
\includegraphics[height=2.5in,width=4.5in, angle=0]{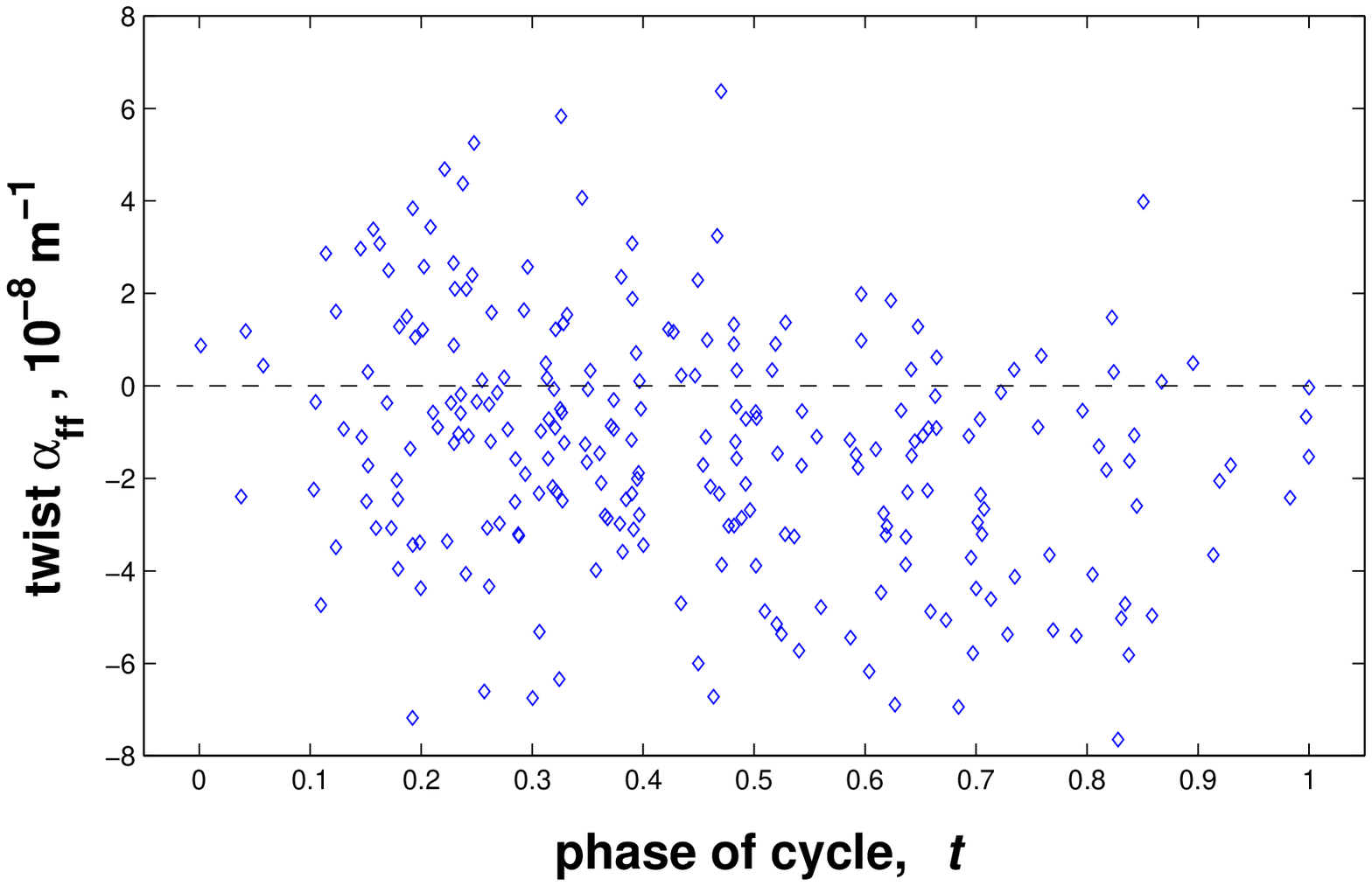}
\vspace{-5mm}
\vspace{-5mm}
\caption{Helicity (upper panel) and twist (lower panel) for
individual active regions as a function of the phase cycle (for the
southern hemisphere). The cycle is not pronounced though one can see
some predominance of negative values in both the helicity
parameters, in accord with the polarity law. } \label{fig3}
\end{figure}

For the helicity and twist data obtained over individual active
regions, the solar cycle is not well-pronounced. As an example, in
Figure~3 we show helicity (upper panel) and twist (lower panel) as
functions of the cycle phase for active regions in the southern
hemisphere. The periodic behaviour is hardly visible in both cases.
Notice that separate presentation of the data for northern and
southern hemisphere is required because of the hemispheric rule
according to which both tracers tend to have opposite signs in the
northern and southern hemispheres.

On the other hand, appropriate averaging of the data taken over
rather narrow temporal or latitudinal intervals indicates similar
cyclic behaviour for both quantities. This is described in the
following section.

\section{Evolution of the Mean Values Over an Activity Cycle
}

The mean values of the twist and helicity calculated over relatively
narrow time or latitudinal intervals behave quite differently.
Following Kleeo\-rin {\it et al.} (2003), we divide the entire data set
into two-year time intervals and plot the mean helicity (Figure~4,
upper panel) and mean twist (Figure~4, lower panel) for each interval
separately for the northern and southern hemisphere. The error boxes
are calculated assuming the quantities are Gaussian. Figure~4 shows
that the cyclic variations of the helicity and twist are very
similar. Both parameters increase in absolute value in the middle of
the cycle and decrease at the beginning and end of the cycle. Both
helicity and twist change sign from one hemisphere to another. For
both tracers, the cycle is seen more distinctly in the southern
hemisphere; the cycle in the northern hemisphere is somewhat better
pronounced in helicity  than in twist. On the other hand, the mean
values contain significant uncertainties. Therefore, we cannot
confirm the hypothesis that the mirror asymmetry of the magnetic
field changes sign in the course of an activity cycle as suggested
by Hagino and Sakurai (2005). We have checked that the result is
robust with respect to discarding data that strongly deviates from
the mean.

\begin{figure}
\includegraphics[height=2.2in,width=4.1in, angle=0]{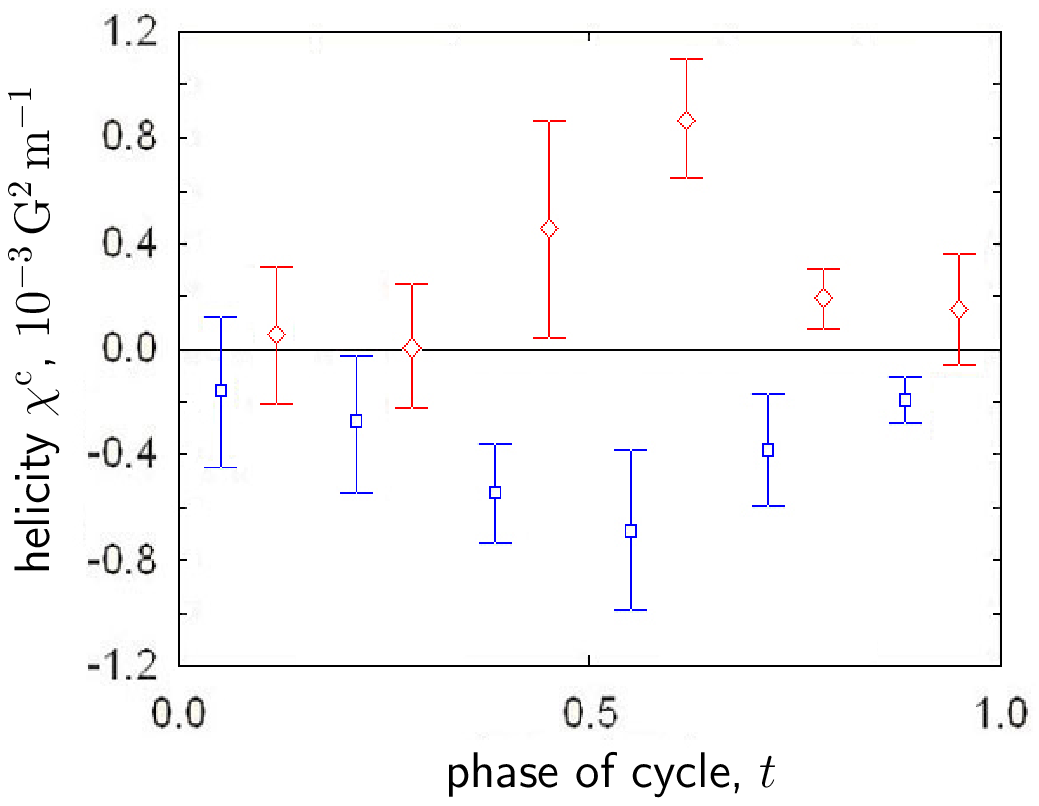}
\vbox{
\vspace{3mm}
\hbox{
\hspace{-3mm}
\includegraphics[height=2.2in,width=4.10in, angle=0]{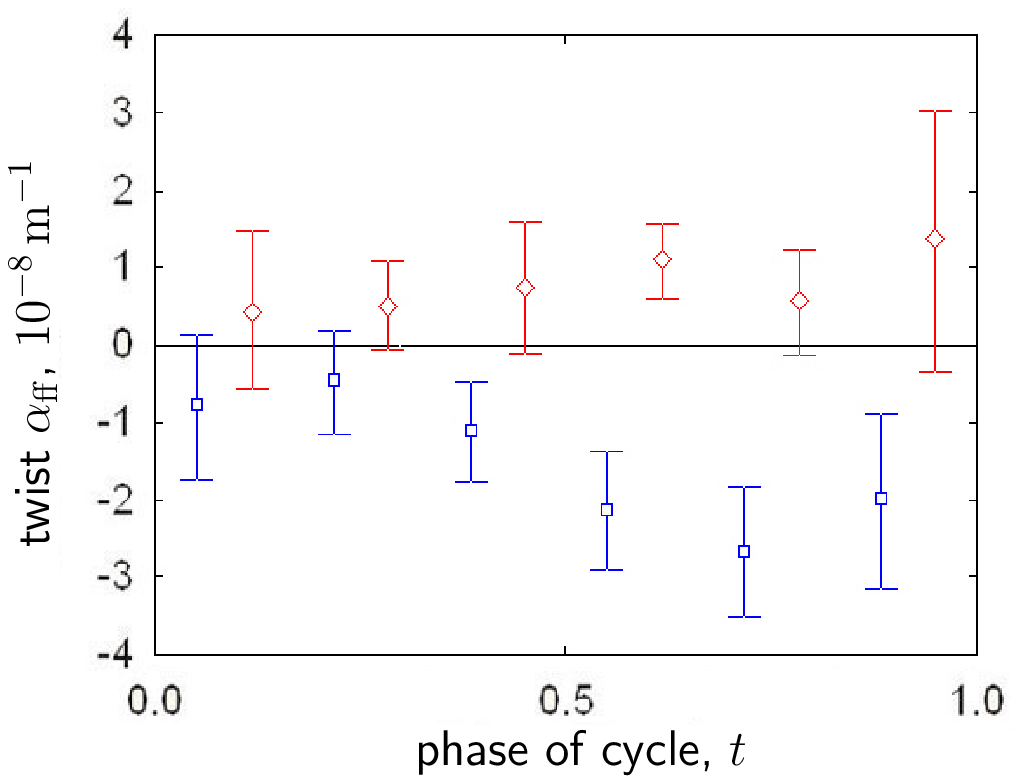}
}
}
\caption{Two-year averaged mean helicity (upper panel) and twist
(lower panel) values as a function of the cycle phase calculated
separately for the northern and southern hemispheres. Blue squares
correspond to the  northern hemisphere while red diamonds correspond
to the southern one. The data points averaged over both hemispheres,
when plotting, have been artificially shifted in opposite sides for
better performance. One can easily see the polarity law and
evolution of both the values over the cycle. } \label{fig5}
\end{figure}

Note that, in comparison with Kleeorin {\it et al.} (2003), we use an
extended database covering a longer period of observations. The
observational database and the procedure providing a synthetic
description of a full cycle using partial observations of two
successive cycles are given in Sokoloff {\it et al.} (2006).

A similar picture was obtained by averaging the data over
$5^\circ$~-latitudinal bands (Figure 5). Again, both tracers display a mirror
asymmetry with respect to the solar equator. The helicity data in
the southern hemisphe\-re are more regular than in the northern one,
while the twist data, on the contrary, seem to be more regular in
the northern hemisphere. Also, we do not see polarity inversions
over latitudinal bands in either hemisphere. (As shown by the
confidence intervals, the apparent polarity reversal at $-30^\circ$
for helicity is insignificant and disappears if strongly deviating
values are discarded). Notice that a similar result was obtained by
Zhang {\it et al.} (2002) using a smaller dataset, see their Figure~4.

\begin{figure}
\includegraphics[height=2.2in,width=4.10in, angle=0]{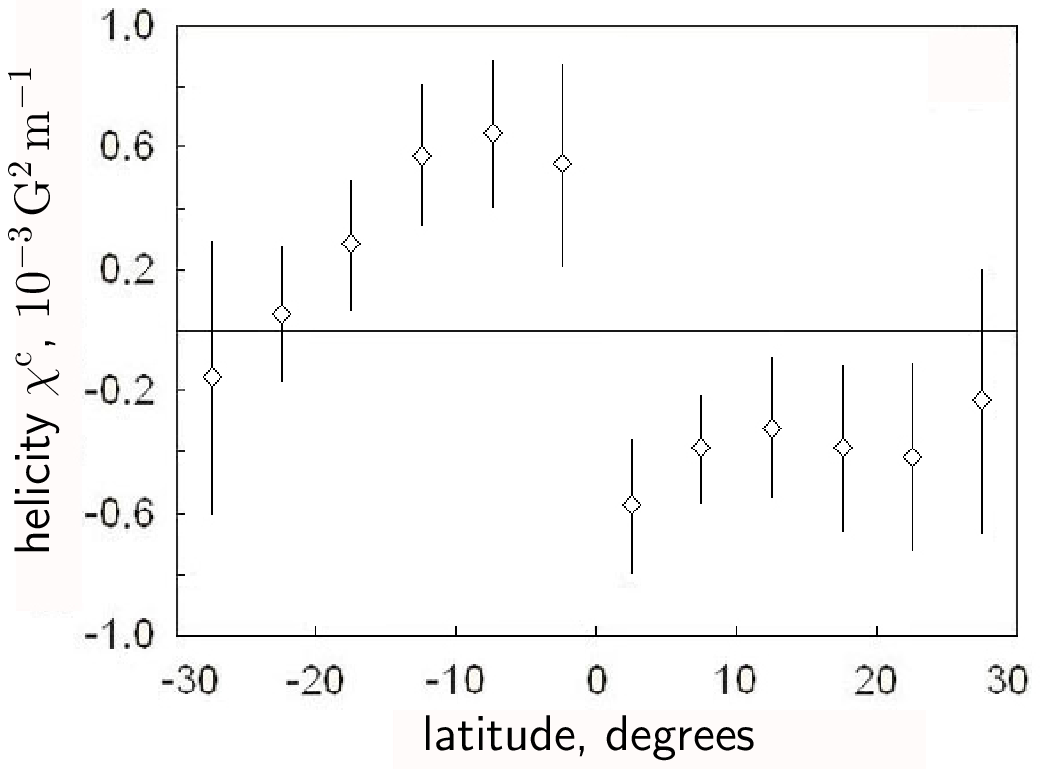}
\vbox{
\vspace{0mm}
\hbox{
\hspace{0mm}
\includegraphics[height=2.2in,width=4.10in, angle=0]{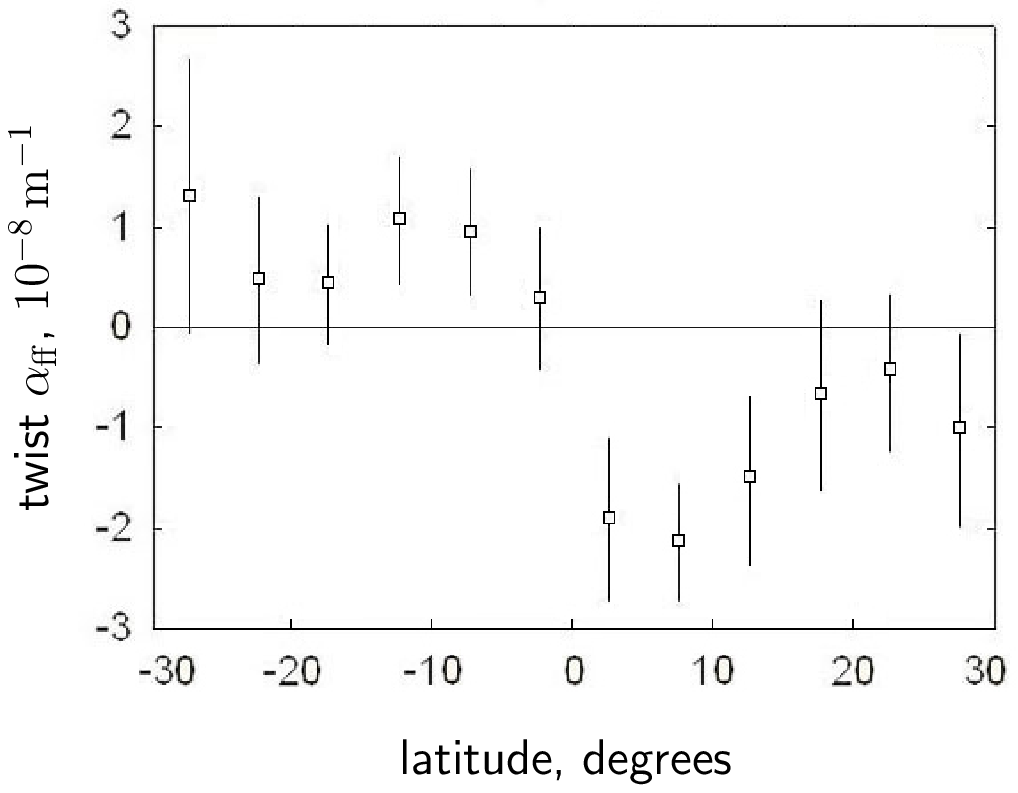}
}
}
\caption{The helicity (upper panel) and twist (lower panel) values
averaged over $5^\circ$~-latitudinal bands.One can easily see the polarity law.
} \label{fig4}
\end{figure}

In general, we conclude that the twist data averaged over time or
latitude intervals can be used to determine quite reliably the
behaviour of the helicity and vice versa.

Isolating the particular time intervals and latitudinal bands and
separat\-ing the data by hemispheres, we decrease significantly the
number of the values to be averaged in each case. Therefore, the
available data appears to be insufficient for
further fragmentation. In particular, butterfly diagrams for the
helicity (Sokoloff {\it et al.}, 2006) and twist (Figure~6) based on these
data are merely illustrative and cannot be used directly to draw
conclusions on, say, the inversion of the sign of helicity.

\begin{figure}
\centering
\includegraphics[height=2.5in,width=4.5in, angle=0]{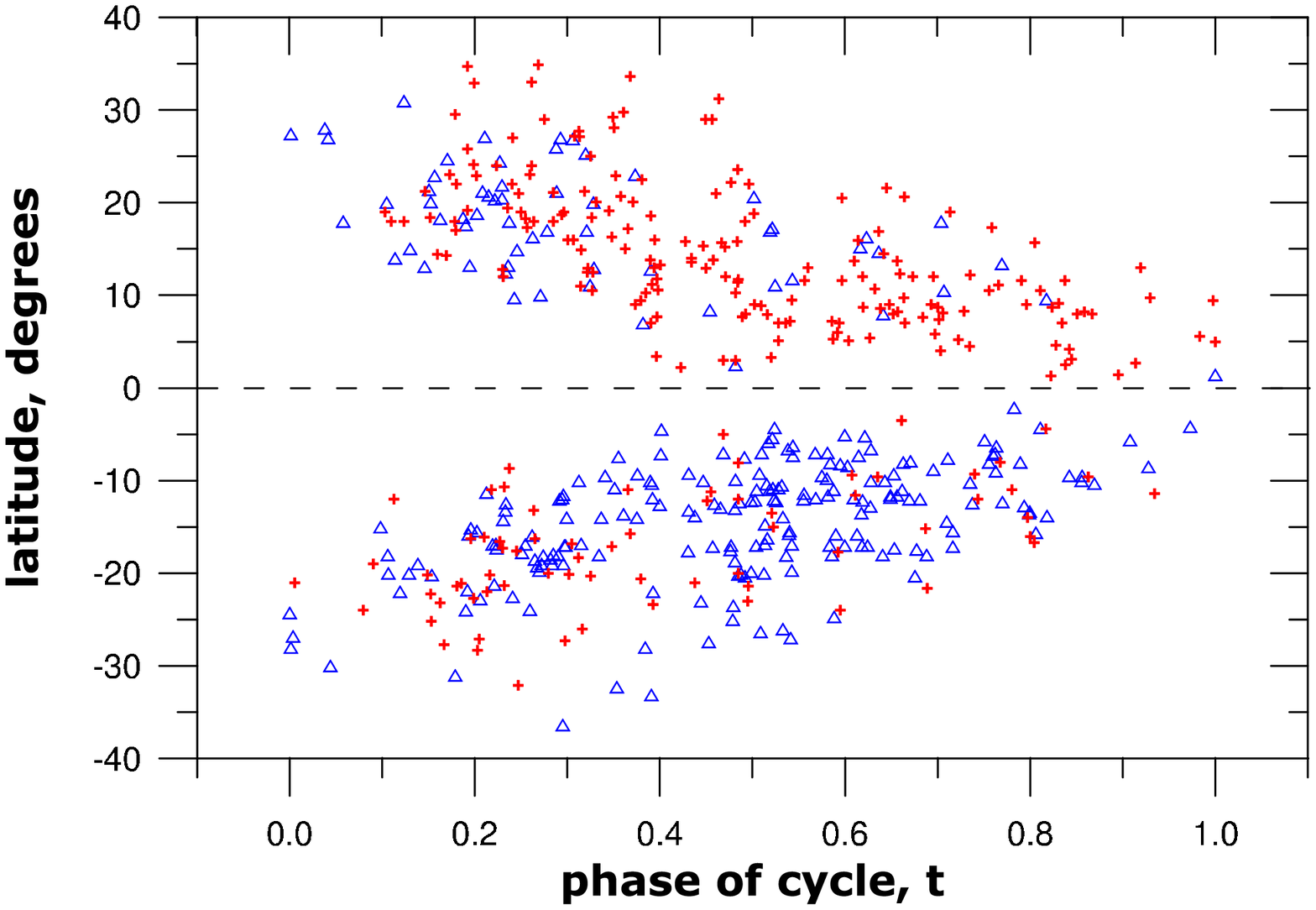}
\includegraphics[height=2.5in,width=4.5in, angle=0]{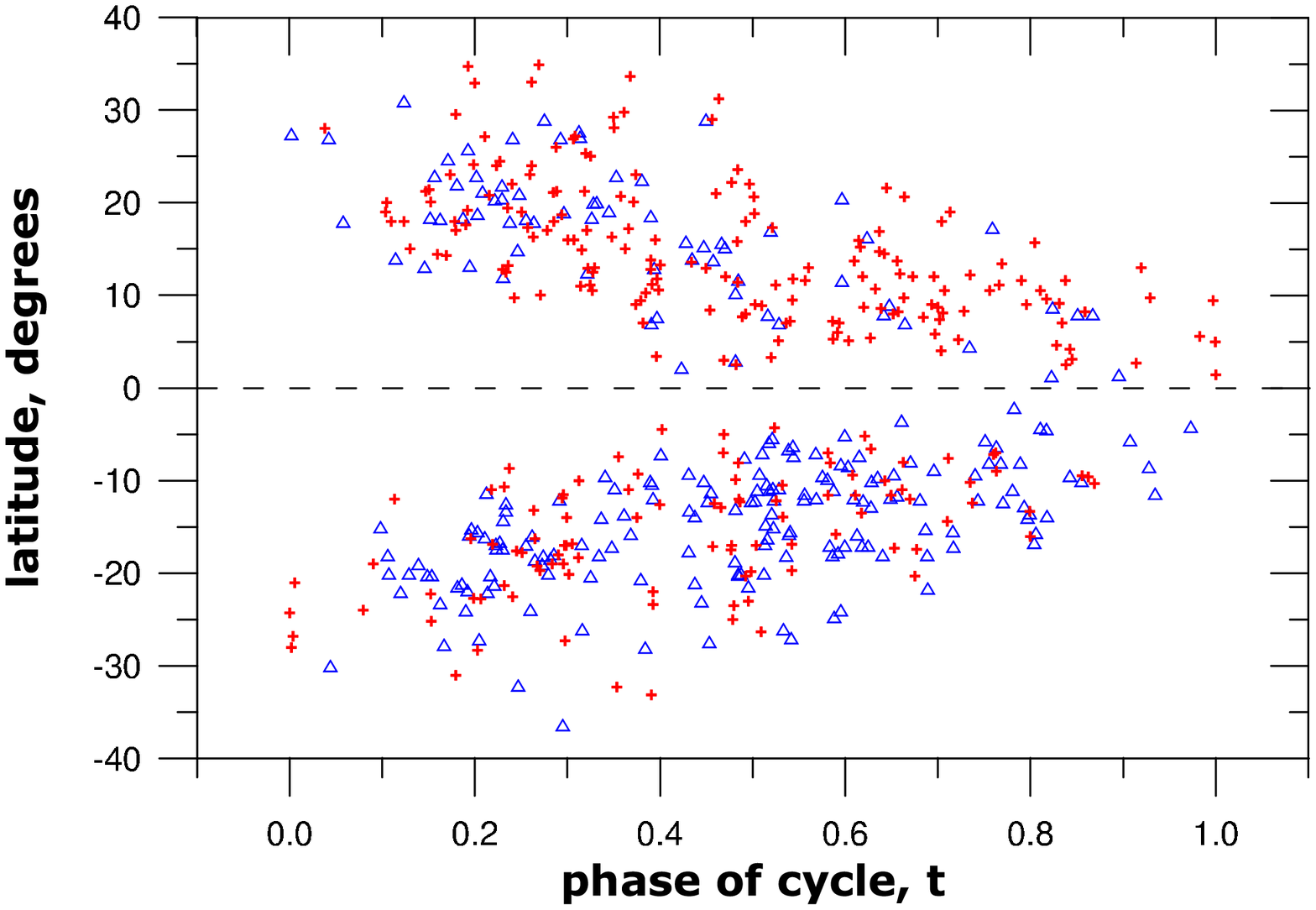}
\caption{
Butterfly diagrams for the
helicity
(upper panel) and
twist
(lower panel).
The red crosses
mark the positions of the active regions
with negative values of the corresponding parameters, and the
blue triangles
correspond to the positive values.
The two diagrams look very
similar, and so can hardly be distinguished by sight.  \label{fig6}}
\end{figure}

Note that a weak correlation between the helicity and twist data
for individual active regions discussed in the previous section
needs an explanation in the context of the revealed pronounced
correlation between their mean values. We may suggest that there {\it is}
significant correlation between the helicity and twist for
individual active regions calculated in a particular latitudinal
range and cycle phase, but this may be disguised by the general cyclic
dependence of the entire dataset.
Thouth the limited size of the available dataset does not allow us
to verify this suggestion.

\section{
Statistics of Active Regions Violating the Polarity Law
}

Of course, the polarity law for the helicity and twist is not
strictly followed. As shown by the helicity and twist measurements
in individual active regions, there are a lot of active regions that
violate this law. It turns out (Sokoloff {\it et al.}, 2006) that the
active regions in which the current helicity does not obey the
polarity law are most frequently observed at the beginning of the
cycle (cf. Tang and Le, 2005). Table~1 presents the corresponding
statistics both for the helicity and twist. The tendency of the
law-breaking active regions to appear preferably at the beginning of
the cycle is noticeable in the case of the twist as well, but this
is much weaker than in the case of the helicity (in contrast to what
might be expected according to Choudhuri {\it et al.} (2004). This raises
a problem which challenges further development of the dynamo theory.

\begin{table}
\begin{tabular}{ccccc}

\hline
$t$ & $n_-$ & $N_-$ & $p$ & $\tilde p$ \\
\hline
0.18 & 18 & 16 & $54\pm 8 \%$ & $48 \pm 18 \%$ \\
0.30 & 60 & 54 & $46\pm 4 \%$ & $42 \pm 9 \%$\\
0.43 & 85 & 91 & $ 37 \pm 3 \%$ &  $40 \pm 6 \%$ \\
0.55 & 101 & 123 & $32 \pm 2 \%$ &  $38 \pm 5 \%$\\
0.68 & 112 & 144 & $28 \pm 2 \%$ &  $ 36 \pm 5 \%$\\
0.80 & 121 & 162 & $26 \pm 2 \%$ &  $35 \pm 4 \%$\\
\hline

\end{tabular}

\caption{
Statistics of active regions breaking the
polarity law for the current helicity ($n_-$ is the number of active
regions before the cycle phase $t$ and $p$ is the relative number
(probability) of the law-breaking active regions) and the twist
($N_-$ is the active region number and $\tilde p$ is the
probability).}
\end{table}

\section{Discussion}

In this paper we have shown that the behaviour of current helicity
and twist is similar. Both these quantities are obtained from the
same distribution of local values of vertical magnetic field $b_z$
and the electric current $j_z$, properly averaged upon calculation.
Therefore this similarity is expected, however not evident in
advance. Although one can suggest other quantities to be calculated
{}from the same vector magnetographic dataset, such as $\alpha_{\rm
best}$, which can also be used as a tracer of mirror asymmetry of
magnetic fields (see, {\it e.g.} Hagino and Sakurai, 2005). We may suggest
to compare these related quantities in the forthcoming papers.

The above analysis exploits sunspots as tracers of processes in the
region of magnetic field generation, {\it i.e.} in the region of dynamo
action. Let us discuss the applicability of this approach. At the
photospheric level magnetic
pressure in sunspots is likely much larger than the gas pressure,
and so the magnetic field can be considered in the vacuum
approximation. This supports the application of potential magnetic
field models for the solar corona and to some extent for the
photosphere. Correspond\-ingly, we use twist as a quantity
reflecting mirror asymmetry of magnetic field in the potential
approximation. The situation becomes quite different in the
sub-photospheric region as the gas pressure and kinetic energy become
comparable with the magnetic energy just below the photosphere.
Correspondingly,
we use current helicity as a quantity
reflecting the mirror asymmetry in the solar interior.

The extent to which the current helicity data taken at the
photos\-pheric level represent values for the dynamo region seems to
be much more delicate. The point is that time-distance
helioseismology  demonstrated that sunspots evanesces at the depth
of 5-10 Mm only. One might conclude that this fact precludes using
sunspots as tracers of physical processes in the solar interior. We
believe that such an opinion would be an exaggeration. First of all,
helioseismology gives the depth of the region with a temperature
depression only. Below this depth a temperature excess is expected
(Ponomarenko, 1972a,b; Parker, 1974, 1976). It is reasonable to
believe that the magnetic field below a sunspot cannot suppress
convection even though it is not negligible. Note that according to
modern understanding in the framework of a cluster model, the motion
of magnetic flux tubes is rather independent of convective motion.
This viewpoint is supported by direct observations (Zhao {\it et al.},
2004; Gizon  et al, 2003). In addition, long term monitoring of
sunspot rotation demonstrates a clear solid-body component typical
for tachocline motion (Ivanov, 2004). One may expect that sunspots
mimic somehow the structure of the large-scale toroidal magnetic
fields in the solar interior. Therefore, we may use the data on
current helicity and twist as tracers of the dynamo mechanism,
though this question requires further clarification.

\begin{acknowledgements}
We would like to thank the anonymous referee for constructive
criticism which helped to improve the paper. The authors (D.S. and
K.K.) would like to acknowledge support from Chinese Aca\-demy of
Sciences and NSFC towards their visits to Beijing under projects
RFBR-NNSFC 05-02-39017 and RFBR  06-05-64619, 05-02-16090, 
07-02-00246 and 08-02-00070, and also Leading Schools grant HIII-8499.2006.2. 
The work was supported by National Basic Research
Program of China 2006CB806301, National Natural Science Foundation
of China 106111\-20338, 10473016, 10673016, 60673158, and Important
Directional Pro\-ject of Chinese Aca\-demy of Sciences KLCX2-YW-T04.
DS is gratefull to the Royal Society for financial support of his visit
to the U.K. We are gratefull to Andrew Fletcher for critical reading of
the manuscript.

\end{acknowledgements}

\end{article}

\begin{thebibliography}{}

\bibitem[\protect\citeauthoryear{Abramenko {\it et al.}}{1996}]{abram96}
Abramenko, V.I., Wang, T.J., Yurchishin, V.B.: 1996, {\it Solar Phys.} {\bf 
}{\bf
168}, 75.

\bibitem[\protect\citeauthoryear{Abramenko {\it et al.}}{1997}]{abram97}
Abramenko, V.I., Wang, T.J., Yurchishin, V.B.: 1995 {\it Solar Phys.
}{\bf
174} 291.

\bibitem[\protect\citeauthoryear{Bao {\it et al.}}{2000}]{bao00}
Bao, S.D., G.X. Ai, H.Q. Zhang, 2000: {\it J. Astrophys. Astron.
}{\bf
21}, 303.

\bibitem[\protect\citeauthoryear{Bao {\it et al.}}{2002}]{bao02}
Bao, S.D., Ai, G.X., Zhang, H.Q.: 2002,
In: Rickman, H. (ed.),
{\it IAU Highlights Astron.} {\bf 12}, 392.

\bibitem[\protect\citeauthoryear{Bao and Zhang}{1998}]{bao98}
Bao, S.D., Zhang, H.Q.: 1998, {\it Astrophys. J. 
}{\bf
496}, L43.

\bibitem[\protect\citeauthoryear{Choudhuri {\it et al.}}{2004}]{chou04}
Choudhuri, A.R., Chatterjee, P., Nandy, D.:  2004, {\it Astrophys. J.
}{\bf
615}, L57.


\bibitem[\protect\citeauthoryear{Gizon {\it et al.}}{2003}]{gizon03}
Gizon, L., Duvall, Jr., T.L., Schou, J.:  2003, {\it Nature
}{\bf
421}, 43.

\bibitem[\protect\citeauthoryear{Hagino and Sakurai}{2005}]{hagino05}
Hagino, M.,  Sakurai, T.: 2005, {\it 
Publ. Astron. Soc. Japan
}{\bf
 57}, 481.


\bibitem[\protect\citeauthoryear{Ivanov}{2006b}]{ivanov06b}
Ivanov, E.V.: 2004,
In: Stepanov, A.V., Benevolenskaya, E.E., Kosovichev, A.G. (eds),
{\it Multi-Wavelength Investigations of Solar Activity, IAU Symp.} {\bf 223},
261.

\bibitem[\protect\citeauthoryear{Kleeorin {\it et al.}}{2003}]{kleeorin03}
Kleeorin, N.,  K. Kuzanyan, D. Moss, I. Rogachevskii, Sokoloff, D.,
Zhang, H.:  2003, {\it Astron. Astrophys.
}{\bf
 409}, 1097.

\bibitem[\protect\citeauthoryear{Kleeorin and Rogachevskii}{1999}]{kleeorin99}
Kleeorin N., Rogachevskii, I.:  1999, {\it Phys. Rev. E
}{\bf
 59}, 6724.

\bibitem[\protect\citeauthoryear{Kuzanyan {\it et al.}}{2000}]{kuzanyan00}
Kuzanyan, K.M., Bao, S.,  Zhang,H.:  2000, {\it Solar Phys.
}{\bf
 191},
231.

\bibitem[\protect\citeauthoryear{Parker}{1955}]{parker55}
Parker, E.N., 1955: {\it Astrophys. J.
}{\bf
 122}, 293.


\bibitem[\protect\citeauthoryear{Parker}{1974}]{parker74}
Parker, E.N.,  1974: {\it Solar Phys.
}{\bf
 36}, 249.

\bibitem[\protect\citeauthoryear{Parker}{1976}]{parker76}
Parker, E.N.,  1976: {\it Astrophys. J.
}{\bf
 204}, 259.

\bibitem[\protect\citeauthoryear{Pevtsov {\it et al.}}{1994}]{pevtsov94}
Pevtsov, A.A.,  R.C. Canfield, T.R. Metchalf,  1994: 
{\it Astrophys. J.
}{\bf
 425}, L117.

\bibitem[\protect\citeauthoryear{Ponomarenko}{1972a}]{ponomarenko72a}
Ponomarenko Yu.B.,  1972a: {\it Astron. Rep.
}{\bf
 49}, 148.

\bibitem[\protect\citeauthoryear{Ponomarenko}{1972b}]{ponomarenko72b}
Ponomarenko Yu.B.,  1972b: {\it Astron. Rep.
}{\bf
 49}, 568.

\bibitem[\protect\citeauthoryear{Rust and Kumar}{1996}]{rust96}
Rust, D.M.,  A. Kumar,  1996: {\it Astrophys. J.
}{\bf
 464}, L119.

\bibitem[\protect\citeauthoryear{Seehafer}{1990}]{seehafer90}
Seehafer, N.,  1990: {\it Solar Phys., 125}, 219--232.

\bibitem[\protect\citeauthoryear{Sokoloff {\it et al.}}{2006}]{sokoloff06}
Sokoloff, D., S.D. Bao, N. Kleeorin, K. Kuzanyan, D. Moss, I.
Rogachevskii, D. Tomin, H. Zhang,  2004: {\it Astron. Nachr.
}{\bf
 327},
876.

\bibitem[\protect\citeauthoryear{Tang and Le}{2005}]{Tang2005}
Tang, Y.Q., Le, G.M.: 2005,
In: Acharya, B.S., Gupta, S., Jagadeesan, P., Jain, A., Karthikeyan, S.,
Morris, S., Tonwar, S. (eds.),
{\it Proc. 29-th International Cosmic Ray Conference},
Tata Institute of Fundamental Research, Vol.~1, 5.

\bibitem[\protect\citeauthoryear{Wang {\it et al.}}{1996}]{wang96}
Wang, T.J., Ai, G.X., Deng, Y.Y.:  1996, {\it Astrophys. Rep. (Publ.
Beijing Obs.)
}{\bf
 28}, 31-32.

\bibitem[\protect\citeauthoryear{Woltjer}{1958}]{woltjer58}
Woltjer, L., 1958: {\it Proc. Natl. Acad. Sci. USA
}{\bf
 44}, 489.

\bibitem[\protect\citeauthoryear{Xu {\it et al.}}{2007}]{xuetal2007}
Xu, H., Gao, Y., Zhang, H., Sakurai, T., Pevtsov, A.A., Sokoloff, D.: 2007,
{\it Adv. Space Res.} {\bf 39}, 1715.

\bibitem[\protect\citeauthoryear{Zhang and Bao}{1998}]{Zhang98}
Zhang, H.,  S. Bao, 1998: {\it  Astron. Astrophys.
}{\bf
 339}, 880.

\bibitem[\protect\citeauthoryear{Zhang {\it et al.}}{2002}]{Zhang02}
Zhang, H.,  S. Bao,  K.M. Kuzanyan, 2002: {\it  Astron. Rep.
}{\bf
 46},
414.

\bibitem[\protect\citeauthoryear{Zhang {\it et al.}}{2006}]{Zhang06}
Zhang, H., D. Sokoloff, I. Rogachevskii, D. Moss, V. Lamburt, K.
Kuzanyan, N. Kleeorin, 2006: {\it  
Monthly Notices Roy. Aston. Soc.
}{\bf
365}, 276.

\bibitem[\protect\citeauthoryear{Zhao {\it et al.}}{2004}]{Zhao04}
Zhao, J., A.G. Kosovichev, T.L. Duvall, 2004: {\it  Astrophys. J.
}{\bf
607}, L135.

\end{thebibliography}
\end{document}